\documentclass[aps,prd,preprint,groupedaddress,showpacs,preprintnumbers]{revtex4-1}
\usepackage{amssymb,amsmath}
\usepackage[dvips]{graphicx}
\usepackage{wick}
\usepackage{color}

\begin{document}

\preprint{KEK-TH-1624}

\title{Effective potential and Goldstone bosons in de Sitter space}

\author{Takashi Arai}
\email[]{araitks@post.kek.jp}

\affiliation{KEK Theory Center, Tsukuba, Ibaraki 305-0801, Japan}


\begin{abstract}
We investigate nonperturbative infrared effects for the $O(N)$ linear sigma model in de Sitter space using the two-particle irreducible effective action at the Hartree truncation level.
This approximation resums the infinite series of so-called superdaisy diagrams.
For the proper treatment of ultraviolet divergences, we first study the renormalization of this approximation
on a general curved background.
Then, we calculate radiatively corrected masses and the effective potential.
As a result, spontaneous symmetry breaking is possible, on the other hand, the Goldstone modes acquire a positive definite mass term due to the screening effects of interaction.
Possible infrared divergence is self-regulated by the mass term.
Furthermore, there is a symmetry restoring phase transition as a function of the Hubble parameter.
In our approximation, the phase transition is of first order.

\end{abstract}

\pacs{04.62.+v, 11.10.Gh, 12.38.Cy}

\maketitle

\section{Introduction}
In de Sitter space, the propagator for a massless minimally coupled scalar field has an infrared divergence~\cite{Allen1}.
This causes a breakdown of perturbative expansion around massless fields.
Moreover, in the theory with a sufficiently small mass, the propagator may have an indefinitely large term.
In this case, perturbation theory can no longer be a valid approximation.
Motivated by the inflationary cosmology, the interpretation and treatment of the infrared divergence has become of interest in recent years~\cite{Woodard1994, de Sitter breaking, Garbrecht}.
Also, it suggests that the strong infrared effects may give novel effects on inflationary physics.

One of the strategies for treating such light scalar fields is resummation of perturbative expansion.
In this method, we include an infinite series of diagrams with certain topology.
Then, it becomes possible to capture some nonperturbative quantum aspects, which cannot be obtained
in ordinary perturbation theory.
In this direction, the $O(N)$ linear sigma model is analyzed by using the large-$N$ expansion in the earlier study~\cite{Serreau}.
Furthermore, in our previous work, we study $\phi^4$ theory using the two-particle irreducible (2PI) effective action~\cite{Arai}.
In both analyses, the fields acquire a positive definite mass term due to nonperturbative screening effects of interaction.
The possible existence of an infrared divergence is self-regulated by the mass term.
On the other hand, these analyses predict different behaviors for the effective potential.
In the large-$N$ analysis, it was shown that the strong infrared effects prevent the existence of spontaneous symmetry breaking.
In contrast, for $\phi^4$ theory, the spontaneous symmetry breaking of $Z_2$ symmetry is possible.
Moreover, the broken $Z_2$ symmetry is restored by the first order phase transition as a function of the Hubble parameter.

The absence of symmetry broken states in de Sitter space is reminiscent of the result for scalar field theory in flat two dimensions.
In fact, the propagator for a massless scalar field in flat two dimensions also has an infrared divergence.
Then, the spontaneous breaking of continuous global symmetry is prohibited, avoiding the presence of massless Goldstone bosons~\cite{Coleman}.
This fact makes the absence of symmetry breaking in de Sitter space plausible.
However, if it is true, there is a discontinuous transition between the broken phase in flat space and the symmetric phase in de Sitter space.
Moreover, it remains open whether the result predicted by the large-$N$ expansion is applicable to any number of $N$.
If a broken phase exists at some $N$, the Goldstone modes appear to cause the infrared divergence.
In such a circumstance, an investigation into whether the construction for well-defined theory requires some infrared cutoff or a natural regularization mechanism like the self-regulation exists is desirable.

In this paper, we study the $O(N)$ linear sigma model by using the 2PI effective action at the Hartree truncation level.
The 2PI effective action enables us to extract some nonperturbative quantum aspects by resumming perturbatively expanded vacuum diagrams~\cite{Jackiw}.
However, it is a nontrivial task to renormalize ultraviolet divergences at a given truncation of the 2PI effective action.
In fact, even at the lowest order truncation of the mean field approximation, renormalization is not trivial, where different renormalization prescriptions predict different results in the literature~\cite{Stevenson, Paz}.
However, recently much progress has been made regarding the renormalization of the 2PI effective action formalism.
In Ref.~\cite{Berges}, a systematic treatment of ultraviolet divergences at a given truncation of the 2PI effective action for scalar field theory in flat space is proposed.
Furthermore, explicit counterterms are constructed for scalar field theory with more complicated symmetry at the 2PI Hartree approximation~\cite{Fejos}.
In our previous study, we derive the counterterms needed for $\phi^4$ theory at the 2PI Hartree truncation on a general curved background~\cite{Arai2}.
In this paper, we extend this renormalization scheme to the $O(N)$ linear sigma model.
Using this renormalization, we investigate the nonperturbative infrared effects on the mass and the effective potential.

There is an earlier study with quite similar analysis, using the mean field approximation~\cite{Prokopec}.
However, we believe that the renormalization in this literature is not appropriate.
This is because ultraviolet divergences are renormalized only by the mass counterterm, which depends on the vacuum expectation value and the Hubble parameter.
Moreover, this inappropriate renormalization obstructs a further study in the framework of the Hartree approximation.
A difference between our study and the preceding work is the renormalization prescription; our renormalization prescription enables us to study in detail at the 2PI Hartree approximation.

This paper is organized as follows.
In Sec. II, we provide detail as to how to renormalize the $O(N)$ model at the Hartree truncation level of the 2PI effective action.
In Sec. III, radiative corrections to the mass and the effective potential are calculated.
Its similarity to finite temperature field theory is mentioned.
Moreover, the consistency between our results and the large-$N$ expansion is discussed.
Sec. IV is devoted to a conclusion and discussion.
In this paper, we use the unit system of $\hbar=c=1$.

\section{Renormalization of the 2PI Hartree approximation for the $O(N)$ linear sigma model}
In this section, we show the renormalization of the 2PI Hartree approximation for the $O(N)$ model in general curved space.
We first renormalize the theory in flat space, then we present the way of extension to curved space.

In flat space, the 2PI effective action for a multi-component scalar field, which is a functional of the vacuum expectation value of the quantum field $v^k$ and the full propagator $G_{ij}$, is given by~\cite{Jackiw}
\begin{equation}
\Gamma[v^k,G_{ij}]=S[v^k]+\frac{i}{2} \log \mathrm{det} [G^{-1}_{ij}]-\frac{i}{2}\mathrm{Tr}[1_{ij}]+\frac{i}{2} \! \int \! \! d^4x  \! \int \! \! d^4 x' G_{0ij}^{-1} [v^k](x,x') G_{ji}(x',x)+\Gamma_2[v^k,G_{ij}],
\end{equation}
where summation is understood in the repeated indices and 
\begin{equation}
i G_{0ij}^{-1}[v^k](x,x')= \frac{\delta^2 S[v^k]}{\delta \phi_i(x) \delta \phi_j(x')}, 
\end{equation}
is an inverse propagator.

$\Gamma_2[v^i,G_{ij}]$ is expressed by $(-i)$ times all of the two-particle irreducible vacuum diagrams with the propagator given by $G$ and vertices given by interaction terms $S_{\mathrm{int}}$ of the shifted action.
$\varphi(x)=\phi(x)-v(x)$ is a shifted field.
Here, a two-particle irreducible diagram is a diagram which cannot be cut in two by cutting only two internal lines, otherwise it is two-particle reducible. Various approximations can be made by truncating the diagrammatic expansion for $\Gamma_2[v,G]$. The mean field and gap equations are given as a stationary condition for $\Gamma[v,G]$ with respect to $v$ and $G$. From these equations, we can solve $G$ as a function of $v$, $G=G[v]$. Then the standard 1PI effective action is obtained by inserting $G[v]$ into $\Gamma[v,G]$, giving $\Gamma_{\mathrm{1PI}}[v]=\Gamma[v,G[v]]$.

For the $O(N)$ linear sigma model with the action
\begin{equation}
S[\phi_i]=-\! \int \! \! d^4x \Bigl[ \frac{1}{2} \phi^i (\Box +m^2) \phi^i+\frac{\lambda}{4N} (\phi^i\phi^i)^2 \Bigr],
\end{equation}
where the index $i$ runs from $1$ to $N$, the shifted action is given by the following
\begin{equation}
\begin{split}
&\!\!\!\! S[\varphi^i+v^i]
=
S[v^i]+\int d^4 x \biggl\{ g^{\mu\nu}\partial_{\mu}\varphi^i\partial_{\nu}v^i-m^2v^i\varphi^i-\frac{\lambda}{4N}4(v^iv^i)(v^j\varphi^j) -\frac{1}{2}\varphi^i(\Box+m^2)\varphi^i \\
& \hspace{3.6cm}-\frac{\lambda}{4N}[2(\varphi^i\varphi^i)(v^jv^j)+4(\varphi^iv^i)^2]
-\frac{\lambda}{4N}[(\varphi^i\varphi^i)^2+4(\varphi^i\varphi^i)(v^j\varphi^j)] \biggr\}.
\end{split}
\end{equation}
From the expression, we can read off the free inverse propagator and the interaction terms of $S_{\mathrm{int}}$ as
\begin{equation}
i G_{0ij}^{-1}(x,x')=-\bigl[ \Box+m^2 \bigr] \delta^{ij}\delta (x-x'),
\label{eq:g0}
\end{equation}
\begin{equation}
S_{\mathrm{int}}[\varphi^i]=-\frac{\lambda}{4N}\! \int \! \! d^4x \Bigl[ 2(\varphi^i\varphi^i)(v^jv^j)+4(\varphi^iv^i)^2\Bigr] 
-\frac{\lambda}{4N}\! \int \! \! d^4x\Bigl[ (\varphi^i\varphi^i)^2+4(\varphi^i\varphi^i)(v^j\varphi^j)\Bigr],
\label{eq:linear sigma model interaction}
\end{equation}
where we have defined $G_{0ij}^{-1}$ to be independent of $v^i$. Furthermore, the diagrams which are constructed from counterterm vertices with only one internal line are considered to be 2PI diagrams.

\begin{figure}
 \includegraphics[width=13cm,clip]{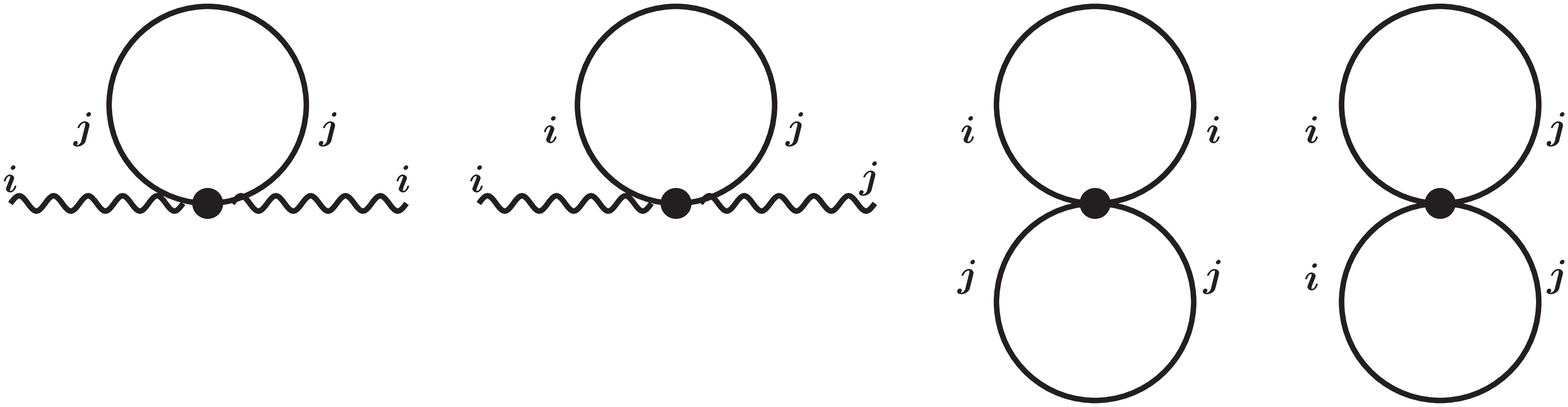}
 \caption{\label{fig:Hartree} 2PI diagrams at the 2PI Hartree approximation. The wiggly line represents the vacuum expectation value of the quantum field, $v^i$.}
 \end{figure}

We now approximate the theory by including only the tadpole and double bubble diagrams as the 2PI diagrams as shown in Fig. \ref{fig:Hartree} and the corresponding counterterm diagrams needed at this truncation. This truncation corresponds to the Hartree-Fock approximation.

\subsection{Renormalization}
In ordinary renormalized perturbation theory, counterterms are perturbatively expanded, and each counterterm vertices in the counterterm diagrams have the same expression at each order in an expansion parameter.
However, in the resummed formalism, in general there are no expansion parameters.
Then, each counterterm vertices in the counterterm diagrams do not necessarily have the same expression.
That is, we have to introduce different counterterm vertices in each counterterm diagram.

In our double-bubble approximation, the reason for introducing different counterterms is explained as the difference in the channels in the resummation process.
At the two-loop order of the 2PI diagrams, there is the double-bubble diagram and the setting sun diagram as shown in Fig. 2.
The double-bubble diagram has the contribution of the diagrams constructed by the $s$-channel.
On the other hand, the setting sun diagram has the contribution of the $t$- and $u$-channel diagrams.
Inclusion of only the double bubble diagrams at the truncation means that we have discarded the contributions of the sub-diagrams constructed by the $t$- and $u$-channels.
How the ultraviolet divergences of the discarded $t$- and $u$-channel diagrams have been engaged to counterterms is different for each counterterm vertex.
Thus, we have to introduce different expressions to each counterterm vertex.

In $\phi^4$ theory, we have to introduce different coupling counterterm vertices $\delta \lambda$ for $S[v]$ and $\Gamma_2[v,G]$.
In the $O(N)$ model, we have to further introduce independent counterterms to the counterterm diagrams of $\Gamma_2[v,G]$ distinguished by the way of contraction.
The reason is explained by the different role of each counterterm vertex diagrams in the renormalization process (see Appendix A).

First, two tadpole diagrams as shown in Fig. \ref{fig:linear sigma model tadpole} can be constructed from the two point interaction terms in Eq. (\ref{eq:linear sigma model interaction}). We assign the following independent counterterms for each contractions
\begin{equation}
\begin{split}
&-\frac{\lambda}{2N}\Bigl[  \wick{1 1}{(v^jv^j)(<2\varphi^i >2\varphi^i)+2(<1\varphi^iv^i)(>1\varphi^jv^j)}  \Bigr], \\
\Rightarrow&-\frac{1}{2N}\Bigl[  \lambda_2^A(v^jv^j) \sum_i G_{ii}+2\lambda_2^B \sum_i (v^iv^i)G_{ii} \Bigr],
\end{split}
\end{equation}
where the superscripts, $A$ and $B$, refer to the different counterterms as $\lambda^A=\lambda+\delta \lambda^A$ and $\lambda^B=\lambda+\delta \lambda^B$, though the finite part is the same.

From the four point interaction terms in Eq. (\ref{eq:linear sigma model interaction}), we can construct the two double bubble diagrams as shown in Fig. 3. Again, we assign the different counterterms as
\begin{equation}
\begin{split}
&-\frac{\lambda}{4N}\Bigl[ 
\wick{1121}{(<1\varphi^i>1\varphi^i)(<2\varphi^j>2\varphi^j)+(<3\varphi^i<4\varphi^i)(>4\varphi^j>3\varphi^j)\times 2} \Bigr], \\
\Rightarrow&-\frac{1}{4N}\Bigl[ \lambda_0^A \sum_i G_{ii} \sum_j G_{jj}+2\lambda_0^B \sum_i G_{ii}^2 \Bigr].
\end{split}
\end{equation}
In the case of four point interaction, we assign the different counterterms by the different ways of contraction, though the interaction term is the same.

 \begin{figure}
 \includegraphics[width=14cm,clip]{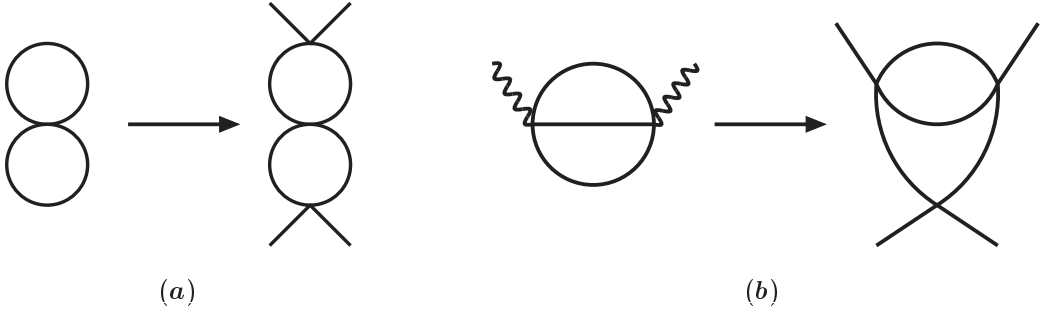}
 \caption{\label{fig:linear sigma model tadpole} 2PI diagrams at the two-loop order: (a) the double bubble diagram and (b) the setting sun diagram. The double bubble diagrams has the contribution of the $s$-channel. On the other hand, the setting sun diagram has the contribution of the $t$- and $u$-channels.}
 \end{figure}

 \begin{figure}
 \includegraphics[width=14cm,clip]{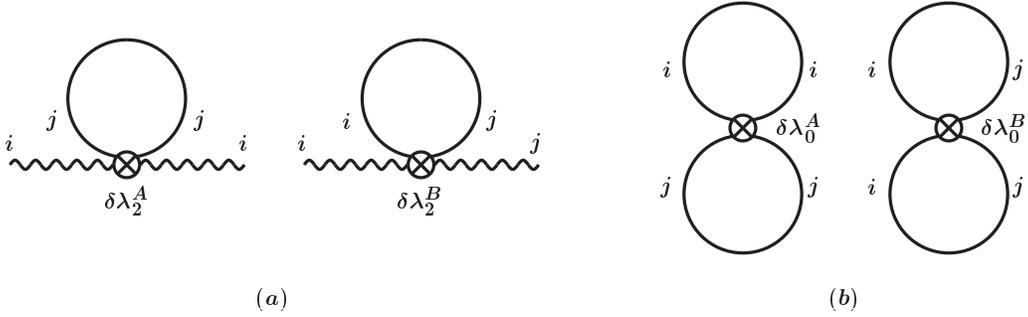}
 \caption{\label{fig:linear sigma model tadpole} The counterterm diagrams of (a) the tadpole consisting of the contraction of the two point interactions, and (b) the double bubble consisting of the contraction of the four point interaction.}
 \end{figure}

With the aid of these independent counterterms, we finally arrive at the following 2PI effective action at the 2PI Hartree approximation in the symmetric phase
\begin{equation}
\begin{split}
\Gamma[v^i,G_{ij}]=&-\int d^4 x \biggl[ \frac{1}{2}v^i(\Box+m_0^2)v^i+\frac{\lambda_4}{4N}(v^iv^i)^2 \biggr]+\frac{i}{2}\ln \det[G_{ij}^{-1}]-\frac{i}{2}\mathrm{Tr}[1_{ij}] \\
&-\frac{1}{2}\mathrm{Tr}\Bigl[ (\Box+m_0^2+\frac{1}{N}\lambda_2^A(  \sum_k v^kv^k)+\frac{2}{N}\lambda_2^B v^iv^i)G_{ij}(x,x')\Bigr] \\
&-\frac{1}{4N}\int \!d^4 x\Bigl[ \lambda_0^A \sum_i G_{ii}(x,x)  \sum_j G_{jj}(x,x)+2\lambda_0^B \sum_i G_{ii}^2(x,x) \Bigr],
\end{split}
\end{equation}
where we have introduce the independent coupling counterterm $\delta \lambda_4$ for $S[v]$ as in the case of $\phi^4$ theory and $m_0^2=m^2+\delta m_0$.

\subsection{Renormalization in the broken phase}
In this paper, we are interested in the broken phase and we renormalize the ultraviolet divergence in the broken phase.
Of course, this renormalization is consistent with that in the symmetric phase.

\begin{figure}
 \includegraphics[width=15cm,clip]{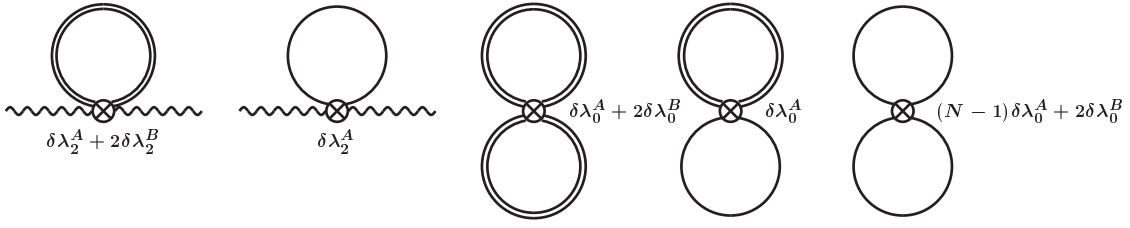}
 \caption{\label{fig:broken bubble} The counterterm diagrams at the Hartree truncation in the broken phase. The double line represents the $\sigma$ full propagator and the single line represents the $\pi^i$ full propagator.}
 \end{figure}

We now break the phase by specifying the vacuum expectation values as $v^1=v$, $v^i=0, i\ne 1$ and rename the fields as $\varphi^1=\sigma$, $\varphi^i=\pi^i, i\ne 1$. In these variables, the 2PI effective action is reexpressed as
\begin{equation}
\begin{split}
\Gamma[v,G_{\sigma},G_{\pi}]=&-\int d^4 x \biggl[ \frac{1}{2}v(\Box+m_0^2)v+\frac{\lambda_4}{4N}v^4 \biggr] +\frac{i}{2}\ln\det[G_{\sigma}^{-1}]+\frac{i}{2}(N-1)\ln\det[G_{\pi}^{-1}] \\
&-\frac{i}{2}N\mathrm{Tr}[1]-\frac{1}{2}\mathrm{Tr}[(\Box+m_0^2+\frac{1}{N}(\lambda_2^A+2\lambda_2^B)v^2)G_{\sigma}] \\
&-\frac{1}{2}(N-1)\mathrm{Tr}[(\Box+m_0^2+\frac{1}{N}\lambda_2^A v^2)G_{\pi}]-\frac{1}{4N}\int \!d^4x (\lambda_0^A+2\lambda_0^B)G_{\sigma}^2 \\
&-\frac{1}{2N}(N-1)\int \! d^4x \lambda_0^A G_{\sigma}G_{\pi}-\frac{1}{4N}(N-1)\int \!d^4x \Bigl[(N-1)\lambda_0^A+2\lambda_0^B\Bigr]G_{\pi}^2,
\end{split}
\end{equation}
where we have identified the various species of the fields $\pi^i$ as $\pi$. The counterterm diagrams contributing to this expression are shown in Fig. \ref{fig:broken bubble}. In the above expression, the different counterterms derived by the different contractions effectively appear as different counterterms for the $\sigma$ and $\pi^i$ fields.

Equations of motion for $v$, $G_{\sigma}$, $G_{\pi}$ are derived by the derivatives of the 2PI effective action with respect to these variables. They are written as follows:
\begin{equation}
-\Bigl[ \Box+m_0^2+\frac{1}{N}\lambda_4 v^2+\frac{1}{N}(\lambda_2^A+2\lambda_2^B)G_{\sigma}+\frac{1}{N}(N-1)\lambda_2^AG_{\pi}\Bigr] v(x)=0,
\end{equation}
\begin{equation}
-\Bigl[ \Box+m_0^2+\frac{1}{N}(\lambda_2^A+2\lambda_2^B)v^2+\frac{1}{N}(\lambda_0^A+2\lambda_0^B)G_{\sigma}+\frac{1}{N}(N-1)\lambda_0^A G_{\pi}\Bigr] G_{\sigma}(x,x')=i\delta(x-x'),
\end{equation}
\begin{equation}
-\Bigl[ \Box+m_0^2+\frac{1}{N}\lambda_2^A v^2+\frac{1}{N}\lambda_0^A G_{\sigma}+\frac{1}{N}\bigl[ (N-1)\lambda_0^A +2\lambda_0^B\bigr] G_{\pi}\Bigr]G_{\pi}(x,x')=i\delta(x-x').
\end{equation}
We first assume that the equations of motion are renormalized by the MS scheme, that is, only the divergent terms are removed by the counterterms, and derive the equations for the counterterms to satisfy. Then, we explicitly construct the counterterms by solving them. From the above equations of motion, we see that one needs $\delta \lambda_0^{A,B}=\delta \lambda_2^{A,B}$ and $\delta \lambda_4=\delta \lambda_2^A+2\delta \lambda_2^B$ for a consistent renormalization. Strictly speaking, the last equality can be derived from the renormalization of the four point vertex function of the $\sigma$ field. This can be seen in Appendix A.

From the equations of motion for the propagators, we define the mass of the $\sigma$ and $\pi^i$ fields as
\begin{equation}
m_{\sigma}^2=m_0^2+\frac{1}{N}(\lambda_2^A+2\lambda_2^B)v^2+\frac{1}{N}(\lambda_0^A+2\lambda_0^B)G_{\sigma}+\frac{1}{N}(N-1)\lambda_0^A G_{\pi},
\end{equation}
\begin{equation}
m_{\pi}^2=m_0^2+\frac{1}{N}\lambda_2^A v^2+\frac{1}{N}\lambda_0^A G_{\sigma}+\frac{1}{N}\bigl[ (N-1)\lambda_0^A +2\lambda_0^B\bigr] G_{\pi},
\end{equation}
and assume that these expressions are renormalized as follows
\begin{equation}
m_{\sigma}^2=m^2+\frac{3}{N}\lambda v^2+\frac{3}{N}\lambda T_F^{\sigma}+\frac{1}{N}(N-1)\lambda T_F^{\pi},
\end{equation}
\begin{equation}
m_{\pi}^2=m^2+\frac{1}{N}\lambda v^2+\frac{1}{N}\lambda T_F^{\sigma}+\frac{1}{N}(N+1)\lambda T_F^{\pi},
\end{equation}
where we have used the divergence structure for tadpole correction, $G_{\sigma,\pi}=m_{\sigma,\pi}^2T_d+T_F^{\sigma,\pi}$. Then, the counterterms have to satisfy the following equations,
\begin{equation}
\begin{split}
&\delta m_0+\frac{1}{N}(\delta \lambda_0^A+2\delta \lambda_0^B)v^2+\frac{1}{N}(3\lambda+\delta \lambda_0^A+\delta\lambda_0^B)m_{\sigma}^2T_d+\frac{1}{N}(\delta \lambda_0^A+2\delta \lambda_0^B)T_F^{\sigma} \\
&\hspace{4.7cm}+\frac{1}{N}(N-1)(\lambda+\delta \lambda_0^A)m_{\pi}^2T_d+\frac{1}{N}(N-1)\delta\lambda_0^A T_F^{\pi}=0,
\end{split}
\end{equation}
\begin{equation}
\begin{split}
&\delta m_0+\frac{1}{N}\delta \lambda_0^A v^2+\frac{1}{N}(\lambda+\delta\lambda_0^A)m_{\sigma}^2T_d+\frac{1}{N}\delta\lambda_0^AT_F^{\sigma} \\
&\hspace{0.7cm}+\frac{1}{N}\bigl[(N+1)\lambda+(N-1)\delta\lambda_0^A+2\delta\lambda_0^B\bigr]m_{\pi}^2T_d
+\frac{1}{N}\bigl[(N-1)\delta\lambda_0^A+2\delta\lambda_0^B\bigr]T_F^{\pi}=0.
\end{split}
\end{equation}
We can explicitly construct the counterterms for the MS scheme by solving these equations. For this, a crucial step is to use the renormalized expressions for the mass. Collecting the terms proportional to $v^2$, $T_F^{\sigma}$ and $T_F^{\pi}$, we obtain
\begin{equation}
\begin{split}
&\delta m_0+\frac{1}{N}(N+2)m^2\lambda T_d+\delta \lambda_0^A m^2T_d+\frac{2}{N}\delta \lambda_0^B m^2T_d \\
&+v^2\frac{1}{N}\biggl[\frac{1}{N}(N+8)\lambda^2T_d+\delta\lambda_0^A+2\delta\lambda_0^B+\frac{1}{N}(N+2)\delta\lambda_0^A \lambda T_d+\frac{6}{N}\delta\lambda_0^B\lambda T_d\biggr] \\
&+T_F^{\sigma}\frac{1}{N}\biggl[\frac{1}{N}(N+8)\lambda^2T_d+\delta\lambda_0^A+2\delta\lambda_0^B+\frac{1}{N}(N+2)\delta\lambda_0^A\lambda T_d+\frac{6}{N}\delta\lambda_0^B \lambda T_d\biggr] \\
&+T_F^{\pi}\frac{1}{N}(N-1)\biggl[\frac{1}{N}(N+4)\lambda^2T_d+\delta \lambda_0^A+\frac{1}{N}(N+2)\delta\lambda_0^A\lambda T_d+\frac{2}{N}\delta\lambda_0^B\lambda T_d\biggr]=0,
\end{split}
\end{equation}
\begin{equation}
\begin{split}
&\delta m_0+\frac{1}{N}(N+2)m^2\lambda T_d+\delta \lambda_0^A m^2T_d+\frac{2}{N}\delta \lambda_0^B m^2T_d \\
&+v^2\frac{1}{N}\biggl[\frac{1}{N}(N+4)\lambda^2T_d+\delta\lambda_0^A+\frac{1}{N}(N+2)\delta\lambda_0^A \lambda T_d+\frac{2}{N}\delta\lambda_0^B\lambda T_d\biggr] \\
&+T_F^{\sigma}\frac{1}{N}\biggl[\frac{1}{N}(N+4)\lambda^2T_d+\delta\lambda_0^A+\frac{1}{N}(N+2)\delta\lambda_0^A\lambda T_d+\frac{2}{N}\delta\lambda_0^B \lambda T_d\biggr] \\
&+T_F^{\pi}\frac{1}{N}\biggl[(N+3)\lambda^2T_d+(N-1)\delta \lambda_0^A+2\delta\lambda_0^B+\frac{1}{N}(N-1)(N+2)\delta\lambda_0^A\lambda T_d \\
&\hspace{9.35cm}+\frac{2}{N}(N+1)\delta\lambda_0^B\lambda T_d\biggr]=0.
\end{split}
\end{equation}
Here, we assume the divergent terms proportional to $v^2$ and $T_F$ as sub-divergence which comes from the divergence of sub-diagrams and the residual divergent terms as overall-divergence.

We then assume that the terms proportional to $v^2$, $T_F^{\sigma}$ and $T_F^{\pi}$ independently vanish. The vertex counterterms $\delta \lambda$ can be determined by the cancellation of the sub-divergences, and the mass counterterm $\delta m_0$ is determined by the cancellation of the overall-divergence. In fact, we obtain the explicit expressions for $\delta \lambda$ from the equations of $T_F^{\sigma}$ and $T_F^{\pi}$
\begin{equation}
\begin{split}
\delta\lambda_0^B=&-\frac{2}{N}\lambda^2T_d\Bigl[1+\frac{2}{N}\lambda T_d\Bigr]^{-1}, \\
=&\lambda \sum_{n=1}^{\infty}\Bigl(-\frac{2}{N}\lambda T_d\Bigr)^n,
\end{split}
\end{equation}
\begin{equation}
\begin{split}
\delta\lambda_0^A=\Bigl[1+\frac{1}{N}(N+2)\lambda T_d\Bigr]^{-1}\Bigl(-\frac{1}{N}\lambda T_d\Bigr)\Bigl[(N+4)\lambda+2\delta\lambda_0^B\Bigr].
\end{split}
\end{equation}
The mass counterterm $\delta m_0$ is constructed from the cancellation of the overall-divergence as
\begin{equation}
\begin{split}
\delta m_0=&-m^2 T_d\biggl[\frac{1}{N}(N+2)\lambda+\delta\lambda_0^A+\frac{2}{N}\delta\lambda_0^B\biggr], \\
=&-m^2\frac{1}{N}(N+2)\lambda T_d\biggl[1+\frac{1}{N}(N+2)\lambda T_d\biggr]^{-1}, \\
=&m^2 \sum_{n=1}^{\infty}\biggl(-\frac{1}{N}(N+2)\lambda T_d\biggr)^n.
\end{split}
\end{equation}
We have renormalized all of the equations of motion. The effective action is renormalized by these counterterms.

When we work on curved space, we have to add the following coupling term to background geometry in the action for a consistent renormalization,
\begin{equation}
S[\phi_i]=-\! \int \! \! d^4x \sqrt{-g} \Bigl[ \frac{1}{2} \phi^i (\Box +m^2+\xi R) \phi^i+\frac{\lambda}{4N} (\phi^i\phi^i)^2 \Bigr],
\end{equation}
where $R$ is the Ricci scalar and $\xi$ is a numerical factor of the conformal parameter.
The alteration in the equations of motion is the coincident propagator.
In a general curved background, it is found by the heat kernel technique that a coincident propagator, which is regularized by the dimensional regularization scheme in four dimensions, has the following general form~\cite{Toms}, 

\begin{equation}
G(x,x)=\bigl[ m^2+(\xi-\xi_c)R \bigr] T_d+T_F,
\end{equation}
where $\xi_c=(d-2)/4(d-1)$ and $T_F$ is an arbitrary tadpole correction.
This alteration gives rise to further overall ultraviolet divergence $(\xi-\xi_c)R T_d$ in the equations of motion.
However, the sub-divergence terms of $v^2$, $T_F^{\sigma}$ and $T_F^{\pi}$ undergo no alterations. 
That is, the sub-divergence terms are renormalized by the same counterterms $\delta \lambda$ as flat space.
The additional overall-divergences are renormalized by the counterterm of the newly introduced conformal parameter $\delta \xi_0$.
It has the following form similar to the mass counterterm, 
\begin{equation}
\begin{split}
\delta \xi_0=& (\xi-\xi_c) T_d\biggl[\frac{1}{N}(N+2)\lambda+\delta\lambda_0^A+\frac{2}{N}\delta\lambda_0^B\biggr], \\
=& (\xi-\xi_c) \frac{1}{N}(N+2)\lambda T_d\biggl[1+\frac{1}{N}(N+2)\lambda T_d\biggr]^{-1}, \\
=&-(\xi-\xi_c) \sum_{n=1}^{\infty}\biggl(-\frac{1}{N}(N+2)\lambda T_d\biggr)^n.
\end{split}
\end{equation}
The effective action is also renormalized by these counterterms and the redefinition of the coupling constants in the gravitational action with curvature squared terms, in a similar way to $\phi^4$ theory~\cite{Arai2}.

\section{Nonperturbative infrared effects in de Sitter space}

Throughout this paper, we work with the standard in-out formalism. In a certain curved background, like de Sitter space, the metric has a time dependence, and its nonequilibrium nature may appear. In such a situation, it is known that the standard in-out formalism is not sufficient, and it is more appropriate to take the Schwinger-Keldysh formalism~\cite{Ramsey}. However, we omit the closed-time path index for the Schwinger-Keldysh formalism, since for our approximation order, at the Hartree truncation level of the 2PI effective action, these in-in and in-out formalisms give the same results.

We use the coordinate system for de Sitter space in terms of comoving spatial coordinates $\mathbf{x}$ and conformal time $-\infty < \eta < 0$ in which the metric takes the form
\begin{equation}
\begin{split}
ds^2&=dt^2-e^{2 H t} d\mathbf{x}^2, \\
         &=a(\eta)^2 (d\eta^2-d\mathbf{x}^2),
\end{split}
\end{equation}
where $a(\eta)=-1/H\eta$ is a scale factor and $H$ is a Hubble parameter constant.

From the renormalization in Sec. II, the renormalized effective action is given by
\begin{equation}
\begin{split}
\Gamma[v,M_{\sigma},M_{\pi}]=&\int d^4 x\sqrt{-g} \biggl[
-\frac{1}{2} v(\Box+m^2)v-\frac{\lambda}{4N} v^4
-\frac{1}{2}\int dM_{\sigma}^2 T_F^{\sigma}-\frac{1}{2}(N-1)\int dM_{\pi}^2T_F^{\pi} \\
&+\frac{3}{4N}\lambda T_F^{\sigma 2}+\frac{1}{2N}(N-1)\lambda T_F^{\sigma}T_F^{\pi}+\frac{1}{4N}(N-1)(N+1) \lambda T_F^{\pi 2}
\biggr].
\end{split}
\label{eq:renormalized effective action}
\end{equation}
Furthermore, the renormalized mean field equation and the self-consistent mass equations are respectively given by
\begin{equation}
\Bigl[ M_{\sigma}^2-\frac{2}{N}\lambda v^2 \Bigr] v=0,
\label{eq:vev}
\end{equation}
\begin{equation}
M_{\sigma}^2=M^2+\frac{3}{N}\lambda v^2+\frac{3}{N}\lambda T_F^{\sigma}+\frac{1}{N}(N-1)\lambda T_F^{\pi},
\end{equation}
\begin{equation}
M_{\pi}^2=M^2+\frac{1}{N}\lambda v^2+\frac{1}{N}\lambda T_F^{\sigma}+\frac{1}{N}(N+1)\lambda T_F^{\pi},
\end{equation}
where we have assumed that $v$ is a constant due to the spacetime symmetry of de Sitter space.
From these equations, we can investigate the possibility of spontaneous symmetry breaking.
To do so, we first assume that $v$ has nonzero solutions and $M_{\sigma}^2-2 \lambda v^2/N=0$.
Using this equality, we eliminate the $v$ dependences in the self-consistent mass equations.
Then, the self-consistent mass equations are given by
\begin{equation}
M_{\sigma}^2=-2 M^2-\frac{6}{N}\lambda T_F^{\sigma}-\frac{2}{N}(N-1) \lambda T_F^{\pi},
\label{eq:broken mass equation1}
\end{equation}
\begin{equation}
M_{\pi}^2=-\frac{2}{N}\lambda T_F^{\sigma}+\frac{2}{N}\lambda T_F^{\pi}.
\label{eq:broken mass equation2}
\end{equation}
The possibility of spontaneous symmetry breaking can be investigated from whether these equations have solutions.
Now, the tadpole correction $T_F$ is expressed as the digamma function (see Appendix B).
An analytic search for solutions requires an approximate expansion for the tadpole corrections $T_F^{\sigma}$ and $T_F^{\pi}$.
In doing so, we first assume that the Goldstone bosons are light compared to the Hubble parameter $H$.
Then, the tadpole correction $T_F^{\pi}$ is expanded as
\begin{equation}
T_F^{\pi}=\frac{H^2}{16\pi^2}\biggl[ \frac{6 H^2}{M_{\pi}^2}+\mathcal{O}((\tfrac{M_{\pi}^2}{H^2})^0) \biggr].
\end{equation}
Next, the expansion for $T_F^{\sigma}$ is possible in two ways. 
One is to assume that $M_{\sigma}$ is heavy compared to $H$.
Then, the tadpole correction $T_F^{\sigma}$ is expanded as
\begin{equation}
T_F^{\sigma}=\frac{H^2}{16\pi^2}\biggl[-\frac{4}{3}+\mathcal{O}(\tfrac{H^2}{M_{\sigma}^2})\biggr].
\end{equation}
where we have set the renormalization scale as $-1+\gamma+\log (\tfrac{M_{\sigma}^2}{4\pi \mu^2}) =0$.
These expansions enable us to solve the self-consistent mass equations as an algebraic equation.
Taking the lowest order of the tadpole expansion, the solutions are respectively given by
\begin{equation}
M_{\sigma}^2=-2M^2+\frac{1}{N}(N+5)\lambda \frac{1}{12\pi^2}H^2
-(N-1)H^2\sqrt{\Bigl(\frac{\lambda}{12 \pi^2 N}\Bigr)^2+\frac{3\lambda}{4\pi^2 N}},
\end{equation}
\begin{equation}
M_{\pi}^2=\frac{\lambda H^2}{12\pi^2 N}+ \sqrt{\Bigl(\frac{\lambda H^2}{12\pi^2 N}\Bigr)^2+\frac{3\lambda H^4}{4\pi^2 N}}.
\label{eq:Goldstone mass}
\end{equation}
These solutions indicate that our approximate expansions, $M_{\pi}^2/H^2 \ll 1$ and $M_{\sigma}^2/H^2 \gg 1$, are justified in the parameter region $\lambda \ll 1$ and $H^2/|M^2| \ll 1$, i.e. weak coupling and weak curvature.
The solutions reproduce the masses with spontaneous symmetry breaking in flat space in the limit $H\rightarrow 0$.
Thus, these solutions correspond to the masses at the minimum of the effective potential.
The tadpole correction $T_F^{\sigma}$ can also be approximated by the small mass expansion.
In this case, the self-consistent masses are given as the solutions of an algebraic equation of the fourth degree at the lowest order of the small mass expansion.
This approximation is discussed in the earlier study~\cite{Prokopec}.

The above analysis means that spontaneous symmetry breaking is possible in de Sitter space.
Also, the mass system has two solutions at some Hubble parameter.
This means that the effective potential has two extrema, and shows the signal of the first order phase transition.
From Eq. (\ref{eq:Goldstone mass}), the mass of the Goldstone bosons cannot be zero.
That is, the infrared divergence in the propagator is self-regulated.

These analytic results are reliable only in the limited range of the parameters.
In seeing the behavior of the solutions in all parameter regions, we have to resort to numerical calculation.
Numerical solutions for the mass system of Eqs. (\ref{eq:broken mass equation1}) and (\ref{eq:broken mass equation2}) are shown in Fig. 5,
where the renormalization scale is set as $-1+\gamma+\log (\tfrac{-2 M^2}{4\pi \mu^2}) =0$.
The solution in the symmetric phase $v=0$ is also plotted.
The figure shows the critical value of the Hubble parameter $H_c$ at which
the broken phase solutions become to disappear.
This indicates the phase transition at $H_c$.
Moreover, the mass system has two solutions at some value of $H$.
This is a signal of the first order phase transition.
The mass of the Goldstone bosons cannot be zero.
These results are consistent with those obtained by the analytical method.

The effective mass term of the Goldstone modes superficially appears to be a violation of Goldstone's theorem. However, we can attribute the mass generation to the lack of time-translational invariance in de Sitter space~\cite{Boyanovsky}. In such a circumstance, spontaneous symmetry breaking does not necessarily imply the {\it massless} Goldstone bosons. Thus, the effective mass generation is compatible with symmetry breaking and Goldstone modes.

 \begin{figure}
 \includegraphics[width=8cm,clip]{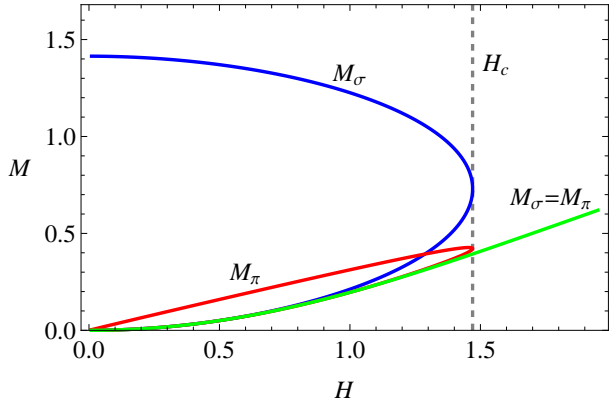}
\caption{\label{fig:linear sigma model mass} The solution of the mass gap system as a function of a Hubble parameter $H$ in the case of $M^2<0$. $N=6$, $\lambda=0.8$ and all values are in the units of $|M|$. The upper and lower curves represent the mass with $v$ on the maximum and minimum respectively of the effective potential as shown in Fig. \ref{fig:linear sigma model potential}. The straight line represents the solution in the symmetric phase.}
 \end{figure}

Here, the following comment should be made.
The solutions of the mass system in Fig. 5 are similar to the same analysis at finite temperature~\cite{Petropoulos}.
In fact, it was found in various contexts that de Sitter space has thermal aspects with the temperature given by $T_H=H/2 \pi$~\cite{Hawking}.
This is due to the existence of the cosmological horizon of de Sitter space.
However, Fig. 5 also shows the difference between finite temperature.
A difference is that the lower curves of the mass solutions always start at $H=0$.
In the terminology of condensed matter physics, the lower spinodal point $H_{c_1}$ always exists at zero value.
On the other hand, in finite temperature field theory, the lower spinodal point exists at nonzero values of temperature.

The nature of the phase transition indicated by the analysis of the mass system can be consistently confirmed by plotting the shape of the effective potentials.
From the renormalized effective action Eq. (\ref{eq:renormalized effective action}), the effective action is given by
\begin{equation}
\begin{split}
V_{\mathrm{eff}}(v,M_{\sigma},M_{\pi})=&
\frac{1}{2} M^2v^2+\frac{\lambda}{4N} v^4
+\frac{1}{2}\int dM_{\sigma}^2 T_F^{\sigma}+\frac{1}{2}(N-1)\int dM_{\pi}^2T_F^{\pi} \\
&-\frac{3}{4N}\lambda T_F^{\sigma 2}-\frac{1}{2N}(N-1)\lambda T_F^{\sigma}T_F^{\pi}-\frac{1}{4N}(N-1)(N+1) \lambda T_F^{\pi 2}.
\end{split}
\end{equation}
Numerical calculation of the effective potentials near the phase transition is shown in Fig. 6, 
where the renormalization scale is set as $-1+\gamma+\log (\tfrac{-2 M^2}{4\pi \mu^2}) =0$ (see Appendix C).
In fact, the first order phase transition takes place at the value of $H_c$ which is indicated by solving the mass system.

 \begin{figure}
 \includegraphics[width=8.7cm,clip]{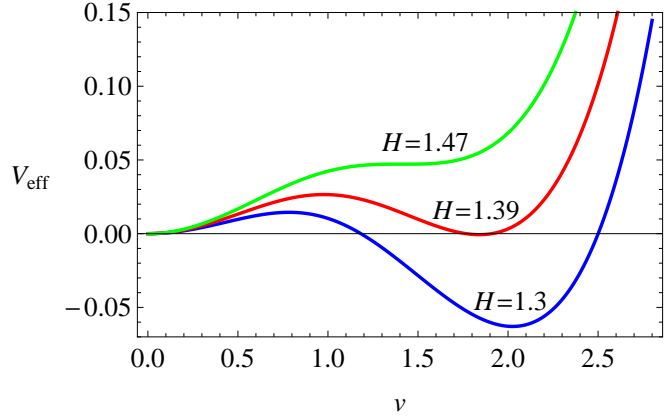}
 \caption{\label{fig:linear sigma model potential} The effective potentials as a function of $v$ near the critical temperature for $N=6$ and $\lambda=0.8$ all in the units of $|M|$. The different lines show the potentials with different values of $H$. Normalization is taken so as to match the origin for different values of $H$.}
 \end{figure}

In the earlier study of the large-$N$ expansion, the strong infrared effects prevent the possible existence of spontaneous symmetry breaking~\cite{Serreau}.
In other words, infrared divergences of the Goldstone modes are circumvented by prohibiting the spontaneously broken phase.
This fact is reminiscent of the absence of the spontaneous breaking of continuous symmetry in flat two dimensions.
In contrast, our result demonstrates that infrared divergences of the Goldstone modes are circumvented by generating the effective mass term of the Goldstone modes rather than the absence of symmetry breaking.
These results appear to be a contradiction.
However, the conflict is reconciled by observing the behavior as $N$ increases in our analysis.
The behavior of the mass solution $M_{\sigma}$ as $N$ increases is shown in Fig. 7.
The figure predicts that the critical value of $H$ decreases as $N$ increases, eventually $H_c$ goes to zero as $N$ goes to infinity.
In fact, if we assume that the small mass expansions of the tadpole corrections $T_F^{\sigma}$ and $T_F^{\pi}$ are valid near the phase transition, the critical value of $H$ can be obtained analytically at large value of $N$. 
At the leading order of $1/N$, it is given by 
\begin{equation}
H_c=2\sqrt{\pi}\biggl(\frac{1}{3\lambda N}\biggr)^{1/4}.
\end{equation}
The validity of this expression is confirmed numerically.
From the expression, we see that $H_c$ becomes zero as $N$ goes to infinity.
This behavior is different from that at finite temperature, there the Goldstone bosons become massless as $N$ goes to infinity.
When the critical curvature is zero, spontaneous symmetry breaking is prohibited and the broken phase in flat space undergoes a discontinuous transition.
Thus, it is concluded that the large-$N$ expansion only tells the nature of the theory as $N$ goes to infinity, predicting the zero critical curvature.

 \begin{figure}
 \includegraphics[width=8.3cm,clip]{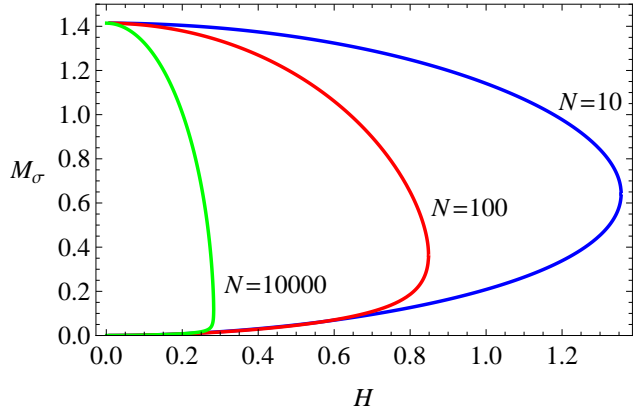}
 \caption{\label{fig:linear sigma model mass large N} The solution of the mass gap system for $M_{\sigma}$ as a function of a Hubble parameter $H$ in the broken phase for various number of $N$. $\lambda=0.8$ and all values are in the units of $|M|$. These solutions indicate that the critical temperature $H_c$ decreases as $N$ increases. It is anticipated that the critical temperature goes to zero in the limit $N\rightarrow \infty$.}
 \end{figure}

\section{Conclusion}

We investigate nonperturbative infrared effects for the $O(N)$ linear sigma model in de Sitter space.
Ultraviolet divergences are regularized by the dimensional regularization scheme.
In all calculations, we work with coordinate space representation.

As a results of our analysis, spontaneous symmetry breaking is possible for any finite number of $N$.
In the broken phase, the possible infrared divergence is circumvented in a manner of generating the mass term for Goldstone modes.
The broken symmetry is restored by the first order phase transition as a function of the Hubble parameter.
These results have a similarity to finite temperature field theory with the temperature given by $T_H=H/2\pi$.
A difference with finite temperature field theory is that the critical temperature has the $N$ dependence of $H_c \sim N^{-1/4}$, predicting a zero value of the critical temperature as $N$ goes to infinity.
From the result, we conclude that the large-$N$ expansion only tells the behavior of the $O(N)$ model as $N$ goes to infinity, i.e. the absence of spontaneous symmetry breaking and the discontinuous transition.
Thus, our results are compatible with those obtained by the large-$N$ expansion.

The effective mass term of the Goldstone modes superficially appears to be a violation of Goldstone's theorem. 
However, we attribute the mass generation to the lack of time-translational invariance in de Sitter space. 
In such a circumstance, spontaneous symmetry breaking does not necessarily imply the {\it massless} Goldstone bosons~\cite{Boyanovsky}. 
Thus, the mass generation is compatible with symmetry breaking and Goldstone modes. 
The effective mass term for Goldstone bosons is also predicted in the same analysis at finite temperature, and this is generally considered as an artifact of the Hartree approximation~\cite{Fejos, Andersen}.
However, we believe that the mass of Goldstone bosons at finite temperature is also caused by the lack of time-translational invariance at the formalism in treating the finite temperature, like the imaginary time formalism~\cite{comment2}.

All our results are based on the mean-field approximation. 
It is necessary to include some contributions of the non-local diagrams, such as the setting-sun type, for a more rigorous study. 
For example, in a similar study at finite temperature, it is found that the order of the phase transition may be changed from the first to the second by the contribution of non-local diagrams~\cite{Oka, Verschelde}. 
This is because in our analysis, it is shown that the results in de Sitter space does not necessarily share the same nature with finite temperature, like its behavior as $N$ varies.
Furthermore, the recent study of the $O(2)$ model by the Wigner-Weisskopf nonperturbative method also predicts the mass generation and first-order phase transition~\cite{Boyanovsky}.

The mass generation of Goldstone bosons is stated in another way that the self-regulation mechanism for IR divergence mentioned in Ref.~\cite{Arai} has taken place in the more complex model. 
We anticipate that the self-regulation mechanism will take place in quantum gravity on a de Sitter background due to field self-interactions.

\appendix
\section{Four-point vertex function of the $\sigma$ field in $O(N)$ model}
In this Appendix, we confirm that the renormalization condition $\delta \lambda_4=\delta \lambda_2^A+2\delta \lambda_2^B$ which comes from the renormalization of the equations of motion is consistent with the renormalization of the four point function of the $\sigma$ field. The four point vertex function of the $\sigma$ field is derived as a fourth derivative of the 1PI effective action as the general form:
\begin{equation}
\begin{split}
\frac{\delta^4 \Gamma_{\mathrm{1PI}}}{\delta v_1 \delta v_2 \delta v_3 \delta v_4}
=&\frac{\delta^4 \Gamma}{\delta v_1 \delta v_2 \delta v_3 \delta v_4}
+\frac{\delta^3 \Gamma}{\delta v_1\delta v_2 \delta G_{ab}^i}\frac{\delta^2 G_{ab}^i}{\delta v_3 \delta v_4}+(6\, \mathrm{perm.}) \\
&+\frac{\delta^2 \Gamma}{\delta G_{ab}^i \delta G_{cd}^j} \frac{\delta^2 G_{ab}^i}{\delta v_1\delta v_2}\frac{\delta^2 G_{cd}^j}{\delta v_3\delta v_4}+(3\, \mathrm{perm.}),
\end{split}
\end{equation}
where we have used a short-hand notation $v_1 \equiv v(x_1)$ and $G_{12} \equiv G(x_1,x_2)$, ``$\mathrm{perm.}$" denotes the possible permutations of the indices 1, 2, 3 and 4, and the repeated indices $i$ and $j$ run from $\sigma$ to $\pi$. Here, we use the relation
\begin{equation}
\begin{split}
\frac{\delta^2 \Gamma}{\delta G_{ab}^i \delta G_{cd}^j}=
\frac{i}{2}\delta^{i \sigma}\delta^{j \sigma} G_{ac}^{\sigma -1}G_{bd}^{\sigma-1}+\frac{i}{2}(N-1) \delta^{i\pi}\delta^{j\pi}G_{ac}^{\pi -1}G_{bd}^{\pi-1}+\frac{\delta^2 \Gamma_2}{\delta G_{ab}^i \delta G_{cd}^j},
\end{split}
\end{equation}
and
\begin{equation}
\frac{\delta^2 G_{ab}^i}{\delta v_1 \delta v_2}=-G_{ac}^{(i)} \frac{\delta^2 G_{cd}^{i -1}}{\delta v_1\delta v_2}G_{db}^{(i)},
\end{equation}
where the index $(i)$ denotes the subsidiary index following to the repeated index $i$.
Then, the four-point vertex function is transformed to
\begin{equation}
\begin{split}
\Gamma_{\mathrm{1PI}}^{(4)}=&
\frac{\delta^4 \Gamma}{\delta v_1\delta v_2 \delta v_3 \delta v_4}
-\frac{\delta^3 \Gamma}{\delta v_1 \delta v_2 \delta G_{ab}^i} G_{ac}^{(i)} \frac{\delta^2 G_{cd}^{i -1}}{\delta v_2 \delta v_4}G_{db}^{(i)}
+(6\, \mathrm{perm.}) \\
&+\frac{i}{2}\frac{\delta^2 G_{bf}^{\sigma -1}}{\delta v_1 \delta v_2} G_{fb}^{\sigma} \frac{\delta^2 G_{dh}^{\sigma -1}}{\delta v_3 \delta v_4} G_{hd}^{\sigma}+\frac{i}{2}(N-1)\frac{\delta^2 G_{bf}^{\pi -1}}{\delta v_1 \delta v_2}G_{fb}^{\pi}\frac{\delta^2 G_{dh}^{\pi -1}}{\delta v_3 \delta v_4}G_{hd}^{\pi}, \\
&+\frac{\delta^2 \Gamma_2}{\delta G_{ab}^i \delta G_{cd}^j} G_{ae}^{(i)}\frac{\delta^2 G_{ef}^{i -1}}{\delta v_1 \delta v_2} G_{fb}^{i} G_{cg}^{(j)} \frac{\delta^2 G_{gh}^{j -1}}{\delta v_3 \delta v_4} G_{hd}^{(j)}+(3\, \mathrm{perm.}).
\end{split}
\end{equation}
We now rewrite the last term by the Bethe-Salpeter equation for $\delta^2 G^{-1}/\delta v^2$. The Bethe-Salpeter equation for the $\sigma$ field is derived as follows
\begin{equation}
\begin{split}
\frac{\delta^2 G_{ab}^{\sigma -1}}{\delta v_1 \delta v_2}=&
\frac{\delta^2}{\delta v_1\delta v_2}\biggl[G_0^{-1}-2i\frac{\delta \Gamma_2}{\delta G_{ab}^{\sigma}}\biggr], \\
=&-2i\biggl[\frac{\delta^3 \Gamma_2}{\delta G_{ab}^{\sigma} \delta v_1\delta v_2}+\frac{\delta^2 \Gamma_2}{\delta G_{ab}^{\sigma}\delta G_{cd}^i}\frac{\delta^2 G_{cd}^i}{\delta v_1\delta v_2}\biggr], \\
=&-2i\biggl[\frac{\delta^3 \Gamma}{\delta G_{ab}^{\sigma}\delta v_1 \delta v_2}-\frac{\delta^2 \Gamma_2}{\delta G_{ab}^{\sigma} \delta G_{cd}^{i}} G_{de}^{(i)}\frac{\delta^2 G_{ef}^{i -1}}{\delta v_1 \delta v_2}G_{fg}^{(i)} \biggr].
\end{split}
\end{equation}
In the same way, the Bethe-Salpeter equation for the $\pi$ field is
\begin{equation}
\begin{split}
\frac{\delta^2 G_{ab}^{\pi -1}}{\delta v_1 \delta v_2}=&
\frac{\delta^2}{\delta v_1\delta v_2}\biggl[G_0^{-1}-\frac{2i}{N-1}\frac{\delta \Gamma_2}{\delta G_{ab}^{\pi}}\biggr], \\
=&-\frac{2i}{N-1}\biggl[\frac{\delta^3 \Gamma}{\delta G_{ab}^{\pi}\delta v_1 \delta v_2}-\frac{\delta^2 \Gamma_2}{\delta G_{ab}^{\pi} \delta G_{cd}^{i}} G_{de}^{(i)}\frac{\delta^2 G_{ef}^{i -1}}{\delta v_1 \delta v_2}G_{fg}^{(i)} \biggr].
\end{split}
\end{equation}
Using the Bethe-Salpeter equation, we obtain
\begin{equation}
\Gamma_{\mathrm{1PI}}^{(4)}=
\frac{\delta^4 \Gamma}{\delta v_1 \delta v_2 \delta v_3 \delta v_4}-
\frac{1}{2}\biggl[ \frac{\delta^3 \Gamma}{\delta v_1 \delta v_2 \delta G_{ab}^i} G_{ac}^{(i)}\frac{\delta^2 G_{cd}^{i -1}}{\delta v_3 \delta v_4} G_{db}^{(i)}+(6\, \mathrm{perm.}) \biggr].
\end{equation}
The last term can also be rewritten by the Bethe-Salpeter-like equation. To see this, we write down the various four-point kernels,
\begin{align}
&\frac{\delta^3 \Gamma}{\delta v \delta v \delta G_{\sigma}}=-\frac{1}{N}(\lambda_2^A+2 \lambda_2^B), \\
&\frac{\delta^3 \Gamma}{\delta v\delta v \delta G_{\pi}}=-\frac{1}{N}(N-1) \lambda_2^A, \\
&\frac{\delta^2 \Gamma_2}{\delta G_{\sigma}\delta G_{\sigma}}=-\frac{1}{2N} (\lambda_0^A+2 \lambda_0^B), \\
&\frac{\delta^2 \Gamma_2}{\delta G_{\sigma}\delta G_{\pi}}=-\frac{1}{2N}(N-1) \lambda_0^A, \\
&\frac{\delta^2 \Gamma_2}{\delta G_{\pi}\delta G_{\pi}}=-\frac{1}{2N}(N-2)\Bigl[(N-1) \lambda_0^A+2 \lambda_0^B\Bigr],
\end{align}
where we have omitted the spacetime indices. We find that the last term is transformed by the Bethe-Salpeter equation for $\delta^2 G_{\sigma}^{-1}/\delta v^2$. Finally, the four point vertex function of the $\sigma$ field is expressed as
\begin{equation}
\begin{split}
\Gamma_{\mathrm{1PI}}^{(4)}=&
\frac{\delta^4 \Gamma}{\delta v_1\delta v_2 \delta v_3 \delta v_4}-\biggl[\frac{\delta^3 \Gamma_2}{\delta G_{12}^{\sigma}\delta v_3 \delta v_4}+\frac{1}{2i}\frac{\delta^2 G_{12}^{\sigma -1}}{\delta v_3 \delta v_4}+(6\, \mathrm{perm.}) \biggr], \\
=&-\frac{6}{N}\lambda+\frac{6}{N}(-\delta \lambda_4+\delta \lambda_2^A+2 \delta \lambda_2^B)
+\frac{1}{2}\Biggl[ \biggl(\frac{\delta^2 G_{12}^{\sigma -1}}{\delta v_3 \delta v_4}+\frac{6}{N}\lambda\biggr)+(6\, \mathrm{perm.})\Biggr].
\end{split}
\end{equation}
From this expression, we see that the condition $\delta \lambda_4=\delta \lambda_2^A+2 \delta \lambda_2^B$ is required for the renormalization of the four point vertex function of the $\sigma$ field. This condition is consistent with the renormalization of the equations of motion.

\section{Coincident propagator}
In this Appendix, we calculate the coincident propagator in de Sitter space to investigate the divergence structure of the tadpole diagram. In $d$ dimensional de Sitter space, the propagator for a free scalar field with mass $m$, conformal factor $\xi$ is expressed by the hypergeometric function~\cite{Candelas}
\begin{equation}
G(x,x')=\frac{H^{d-2}}{(4\pi)^{d/2}}\frac{\Gamma(\frac{d-1}{2}+\nu) \Gamma(\frac{d-1}{2}-\nu)}{\Gamma(\frac{d}{2})} {}_2\mathrm{F}_1\left[\tfrac{d-1}{2}+\nu,\tfrac{d-1}{2}-\nu,\tfrac{d}{2};1+\tfrac{y}{4}\right], 
\end{equation}
where $\nu=\bigl\{ [(d-1)/2 ]^2-(m^2+\xi R)/H^2 \bigr\}^{1/2}$, $R=d (d-1) H^2$ is the Ricci scalar curvature and $y(x,x')=\bigl[ (\eta-\eta')^2-|\mathbf{x}-\mathbf{x}'|^2 \bigr]/\eta \eta '$ is the de Sitter invariant length. In the coincident limit, $y=0$, the formula of the hypergeometric function, ${}_2 \mathrm{F}_1(a,b,c;1)=\Gamma(c)\Gamma(c-a-b)/\bigl[\Gamma(c-a)\Gamma(c-b)\bigr]$, leads to
\begin{equation}
\begin{split}
G(x,x)&=\frac{H^{d-2}}{(4\pi)^{d/2}} \Gamma(1-\tfrac{d}{2})\frac{\Gamma(\frac{d-1}{2}+\nu)\Gamma(\frac{d-1}{2}-\nu)}{\Gamma(\frac{1}{2}+\nu) \Gamma(\frac{1}{2}-\nu)},  \\
& \equiv \frac{H^{d-2}}{(4 \pi)^{d/2}} \Gamma(1-\tfrac{d}{2}) \Gamma(x,x).
\end{split}
\end{equation}
The first gamma function has an ultraviolet divergent pole. The residual gamma function, $\Gamma(x,x)$ determines a coefficient of the ultraviolet divergent pole. 

We restrict our attention to four dimensional spacetime with a regularization parameter $\epsilon=4-d$.
In this case, we can transform the expression $\Gamma(x,x)$ as follows:
\begin{equation}
\begin{split}
\Gamma(x,x)
&=\frac{\Gamma(1+\frac{d-3}{2}+\nu) \Gamma(1+\frac{d-3}{2}-\nu)}
             {\Gamma(\frac{1}{2}+\nu) \Gamma(\frac{1}{2}-\nu)}, \\
&=\Bigl( \frac{d-3}{2}+\nu \Bigr) \Bigl( \frac{d-3}{2}-\nu \Bigr)
      \frac{\Gamma(\frac{d-3}{2}+\nu) \Gamma(\frac{d-3}{2}-\nu)}
             {\Gamma(\frac{1}{2}+\nu) \Gamma(\frac{1}{2}-\nu)}, \\
&=\biggl( \Bigl( \frac{d-3}{2} \Bigr)^2- \Bigl( \frac{d-1}{2} \Bigr)^2+\frac{m^2+\xi R}{H^2} \biggr) \\
      & \ \ \ \ \frac{
            \Gamma(\frac{1}{2}+\nu) \bigl[1+\psi(\frac{1}{2}+\nu)(-\frac{\epsilon}{2})+\mathcal{O}(\epsilon^2) \bigr]
             \Gamma(\frac{1}{2}-\nu) \bigl[1+\psi(\frac{1}{2}-\nu)(-\frac{\epsilon}{2})+\mathcal{O}(\epsilon^2) \bigr]
            }
            {
            \Gamma(\frac{1}{2}+\nu) 
            \Gamma(\frac{1}{2}-\nu)
            },  \\
&=\Bigl( \frac{m^2+\xi R}{H^2}-(d-2) \Bigr) \biggl[1-\Bigl(\frac{\epsilon}{2}\Bigr) \Bigl( \psi(\tfrac{1}{2}+\nu)+\psi(\tfrac{1}{2}-\nu) \Bigr)+\mathcal{O}(\epsilon^2) \biggr], \\
&=\frac{1}{H^2} \left \{ \Bigl( m^2+(\xi - \xi_c) R \Bigr) \biggl[1-\Bigl(\frac{\epsilon}{2}\Bigr) \Bigl( \psi(\tfrac{1}{2}+\nu)+\psi(\tfrac{1}{2}-\nu) \Bigr)+\mathcal{O}(\epsilon^2) \biggr] -\frac{\epsilon}{4} \xi_c R \right\},
\end{split}
\end{equation}
where $\psi(x)$ is the digamma function and $\xi_c = (d-2)/4(d-1)$.
Therefore, the coincident propagator is generally given by
\begin{equation}
\begin{split}
G(x,x)=&\frac{H^2}{16 \pi^2} \Bigl(1-\Bigl( \frac{\epsilon}{2} \Bigr) \log \frac{H^2}{4\pi}+\mathcal{O}(\epsilon^2) \Bigr) \Bigl( -\frac{2}{\epsilon}-1+\gamma+\mathcal{O}(\epsilon) \Bigr) \\
&\frac{1}{H^2} \left \{ \Bigl( m^2+(\xi - \xi_c) R \Bigr) \biggl[1-\Bigl(\frac{\epsilon}{2}\Bigr) \Bigl( \psi(\tfrac{1}{2}+\nu)+\psi(\tfrac{1}{2}-\nu) \Bigr)+\mathcal{O}(\epsilon^2) \biggr] -\frac{\epsilon}{4} \xi_c R \right\}, \\
=&\frac{1}{16 \pi^2} \Bigl( -\frac{2}{\epsilon}-1+\gamma+\mathcal{O}(\epsilon) \Bigr) \\
&\left \{ \Bigl( m^2+(\xi - \xi_c) R \Bigr) \Bigl[1-\Bigl(\frac{\epsilon}{2}\Bigr) \Bigl( \psi(\tfrac{1}{2}+\nu)+\psi(\tfrac{1}{2}-\nu)+\log\frac{H^2}{4\pi \mu^2} \Bigr)+\mathcal{O}(\epsilon^2) \Bigr] -\frac{\epsilon}{4} \xi_c R \right\}, \\
=&\frac{1}{16 \pi^2} \bigl[ m^2+(\xi-\xi_c) R \bigr] \biggl(-\frac{2}{\epsilon} +\psi(\tfrac{1}{2}+\nu)+\psi(\tfrac{1}{2}-\nu)-1+\gamma+\log\frac{H^2}{4\pi \mu^2} \biggr) \\
&+\frac{1}{32\pi^2}\xi_c R +\mathcal{O}(\epsilon), \\
=&\frac{1}{16 \pi^2} \bigl[ m^2+(\xi-\xi_c) R \bigr] \biggl(-\frac{2}{\epsilon} +\psi(\tfrac{1}{2}+\nu)+\psi(\tfrac{1}{2}-\nu)+\log \frac{H^2}{M^2}-1+\gamma+\log\frac{M^2}{4\pi \mu^2} \biggr) \\
&+\frac{1}{32\pi^2}\xi_c R +\mathcal{O}(\epsilon), 
\end{split}
\end{equation}
where $\gamma$ is the Euler-Mascheroni constant.
The renormalization scale $\mu$ is introduced in advance.
Strictly speaking, it is introduced when the coupling constant of the field interaction is made dimensionless in the dimensional regularization scheme.
The above ultraviolet divergence structure appears in any curved background.
In fact, it is revealed by the heat kernel technique that the coincident propagator in an arbitrary background has the form  
\begin{equation}
\begin{split}
G(x,x)=\bigl[ m^2+(\xi-\xi_c)R \bigr] T_d+T_F,
\end{split}
\label{eq:tadpole}
\end{equation}
where $T_d=-2/16\pi^2 \epsilon$ and $T_F$ is a general finite tadpole correction.
That is, the finite tadpole correction $T_F$ in de Sitter space is given by
\begin{equation}
T_F=\frac{1}{16 \pi^2} \bigl[ m^2+(\xi-\xi_c) R \bigr] \biggl(\psi(\tfrac{1}{2}+\nu)+\psi(\tfrac{1}{2}-\nu)-1+\gamma+\log \frac{H^2}{M^2}+\log\frac{M^2}{4\pi \mu^2} \biggr)+\frac{1}{32\pi^2}\xi_c R. 
\end{equation}
Note that the above expressions correctly reproduce the flat space results in the limit $H\rightarrow 0$. 
In the flat space limit, canceling the $H$ dependent terms, the coincident propagator is given by 
\begin{equation}
G(x,x)_{\mathrm{flat}}=\frac{m^2}{16\pi^2}\biggl(-\frac{2}{\epsilon}-1+\gamma+\log \frac{m^2}{4\pi \mu^2}\biggr).
\end{equation}

In performing various calculations, we have to evaluate the tadpole correction $T_F$.
The digamma function in the tadpole correction can be approximately expanded regarding the magnitude of the mass term $M^2$.
First, in the small mass case $M^2/H^2 \ll 1$, $T_F$ is expanded as
\begin{equation}
T_F=\frac{H^2}{16\pi^2}\biggl[ \frac{6 H^2}{M^2}-\frac{20}{3}+4 \gamma-2 \Bigl(-1+\gamma+\log\frac{H^2}{4 \pi \mu^2} \Bigr)+\mathcal{O}(\tfrac{M^2}{H^2})\biggr].
\end{equation}
On the other hand, in the large mass case $M^2/H^2 \gg 1$, $T_F$ is given by
\begin{equation}
T_F=\frac{M^2}{16\pi^2}\biggl[ \Bigl(-1+\gamma+\log \frac{M^2}{4 \pi \mu^2} \Bigr)-\frac{4H^2}{3M^2}
-2 \Bigl(-1+\gamma+\log \frac{M^2}{4 \pi \mu^2} \Bigr) \frac{H^2}{M^2}+\mathcal{O}((\tfrac{H^2}{M^2})^2) \biggr].
\end{equation}

\section{One-loop effective action}
In this Appendix, we calculate the finite part of the one-loop effective action.
This is expressed as the mass integral of the tadpole correction $T_F$,
\begin{equation}
\Gamma^{\mathrm{ren}}_{\mbox{{\scriptsize 1-loop}}}=-\frac{1}{2}\int d^4x \sqrt{-g} \int dM^2 T_F(M^2).
\end{equation}
Using the result in Appendix B, we obtain
\begin{equation}
\begin{split}
&\int dM^2 T_F(M^2) \\ =&\int dM^2\biggl\{ \frac{1}{16 \pi^2} (M^2-2 H^2)\bigl[\psi(\tfrac{1}{2}+\nu)+\psi(\tfrac{1}{2}-\nu)
-1+\gamma+\log\frac{H^2}{4\pi \mu^2} \bigr]+\frac{H^2}{16\pi^2}\biggr\}, \\
=& \frac{H^4}{16\pi^2} \int \frac{dM^2}{H^2}( \frac{M^2}{H^2}-2 )\bigl[ \psi(\tfrac{1}{2}+\nu)+\psi(\tfrac{1}{2}-\nu) \bigr] \\
&+\frac{1}{16\pi^2} \Bigl(\frac{1}{2} M^4-2 H^2 M^2 \Bigr)( -1+\gamma+\log \frac{H^2}{4\pi \mu^2}) +\frac{H^2M^2}{16\pi^2},
\end{split}
\end{equation}
where $\mathcal{O}(\epsilon)$ terms are omitted.
Also, renormalization scale $\mu$ is introduced in advance.
The integral of the first term can be performed by changing the integral variable $M^2$ to $\nu=(9/4-M^2/H^2)^{1/2}$, as follows
\begin{equation}
\begin{split}
&\!\!\!\! \int d\nu \nu (\frac{1}{4}-\nu^2)\bigl[\psi(\tfrac{1}{2}+\nu)+\psi(\tfrac{1}{2}-\nu)\bigr] \\
=&\nu (\frac{1}{4}- \nu^2)\bigl[\log \Gamma(\tfrac{1}{2}+\nu)-\log \Gamma(\tfrac{1}{2}-\nu)\bigr]
-(\frac{1}{4}-3\nu^2)\bigl[\psi^{(-2)}(\tfrac{1}{2}+\nu)+\psi^{(-2)}(\tfrac{1}{2}-\nu)\bigr] \\
&-6\nu\bigl[\psi^{(-3)}(\tfrac{1}{2}+\nu)-\psi^{(-3)}(\tfrac{1}{2}-\nu)\bigr]
+6\bigl[\psi^{(-4)}(\tfrac{1}{2}+\nu)+\psi^{(-4)}(\tfrac{1}{2}-\nu)\bigr].
\end{split}
\end{equation}
Finally, the finite part of the one-loop effective action is given by
\begin{equation}
\begin{split}
\int dM^2 T_F =&
-\frac{H^4}{8\pi^2}\Bigl\{ \nu (\frac{1}{4}-\nu^2)\Bigl[\log\Gamma(\tfrac{1}{2}+\nu)-\log\Gamma(\tfrac{1}{2}-\nu)\Bigr] \\
&-(\frac{1}{4}-3\nu^2)\bigl[\psi^{(-2)}(\tfrac{1}{2}+\nu)+\psi^{(-2)}(\tfrac{1}{2}-\nu)\bigr]  \\
&-6\nu\bigl[\psi^{(-3)}(\tfrac{1}{2}+\nu)-\psi^{(-3)}(\tfrac{1}{2}-\nu)\bigr]
+6\bigl[\psi^{(-4)}(\tfrac{1}{2}+\nu)+\psi^{(-4)}(\tfrac{1}{2}-\nu)\bigr] \biggr\} \\
&+\frac{1}{16\pi^2} \Bigl(\frac{1}{2} M^4-2 H^2 M^2 \Bigr)( -1+\gamma+\log \frac{H^2}{4\pi \mu^2}) +\frac{H^2M^2}{16\pi^2}.
\end{split}
\end{equation}
The above expression has the correct flat space limit as $H$ goes to zero:
\begin{equation}
\begin{split}
\int dM^2T_F(M^2)_{\mathrm{flat}}=&\int dm^2 \frac{m^2}{16\pi^2} \Bigl(-1+\gamma+\log \frac{m^2}{4\pi \mu^2} \Bigr), \\
=&\frac{m^4}{32\pi^2}\Bigl(-1+\gamma+\log \frac{m^2}{4\pi\mu^2}-\frac{1}{2} \Bigr).
\end{split}
\end{equation}

\bibliography{basename of .bib file}

\begin{thebibliography}{99}
\bibitem{comment2}
   First note that the original Goldstone's theorem does not refer to curved space or finite temperature, and thus does not guarantee massless particles.
\end{thebibliography}

\end{document}